\documentclass{aa}
\usepackage{txfonts}
\usepackage{graphicx}
\begin{document}
   \title{Selection criteria for targets of asteroseismic campaigns}

   \author{F.P. Pijpers
          \inst{1}
          }


   \institute{Theoretical Astrophysics Center (TAC),
              Department of Physics and Astronomy (IFA), Aarhus University,
              Ny Munkegade, 8000 \AA{}rhus C, Denmark\\
              \email{fpp@phys.au.dk}
             }

   \date{Received November 11 2002; accepted December 10 2002}

   \abstract{Various dedicated satellite projects are
             underway or in advanced stages of planning to perform 
             high-precision, long duration time series photometry of
             stars, with the purpose of using the frequencies of
             stellar oscillations to put new constraints on the internal
             structure of stars. It is known (cf. \cite{Bro+94}) that
             the effectiveness of oscillation frequencies in
             constraining stellar model parameters is significantly
             higher if classical parameters such as effective
             temperature, and luminosity are known with high
             precision. In order to optimize asteroseismic campaigns
             it is therefore useful to select targets from among
             candidates for which good spectroscopic and astrometric
             data already exists. This paper presents selection
             criteria, as well as redeterminations of stellar
             luminosity and reddening for stars satisfying these
             criteria.
 
   }

   \maketitle
%

\section{Introduction}
The aim of asteroseismology is to use the properties of the
oscillations of stars to place constraints on their internal
structure. The frequencies of stellar oscillations can be used to
determine parameters of stars that are either inaccessible to
classical observation or can not be determined with an accuracy that
is sufficient for many purposes. The foremost examples of these
parameters are the mass and angular momentum of stars. Currently,
accurate masses of stars can only be determined if they are members of
close binary systems. Even for the best spectroscopic data and stellar
atmospheres modelling currently available, for single stars their mass
is not known to better than 20 \%. If one wishes to determine the mass
function of stars and investigate its universality, for instance in
order to investigate star formation processes, the current data are
heavily biased to stars that have formed in binary systems, which may
not reflect the variety of conditions under which stars can form and
thus provide an incorrect mass function for the galaxy. As a
consequence of the fact that forming stars contract from molecular
clouds many orders of magnitude larger than their main sequence
stellar radius, a substantial amount of angular momentum must be shed
during the contraction. An important constraint for theories of
mechanisms mediating this angular momentum loss would be the
measurement of the angular momentum of stars for a range of
masses. Also, for the post-main sequence evolution of in particular
high-mass stars the loss of angular momentum during evolution is
similarly important. Unfortunately spectroscopic observations have
access to only the projected surface rotation velocity. The true
angular momentum can only be obtained through asteroseismology (\cite{Pijp03}).

Oscillation frequencies are measures of sound travel time through the
star, which is an average of the inverse of the sound speed. There is
a relation between the sound speed and the mass because stars are in
hydrostatic equilibrium. There is a secondary dependence on hydrogen
abundance which arises through the change in the mean atomic mass on
which the sound speed depends. Thus if a sufficient number of modes is
present and identified in the time series of an oscillating star, it
should in principle be possible to measure its mass, its core hydrogen
abundance (`age'), and its rotation rate. Using more sophisticated
inverse analyses other information, such as depths of convection
zones, might be possible to extract as well. The identification of
observed modes of oscillation and the subsequent analysis is
considerably facilitated if `external' constraints such as a stellar
effective temperature or a stellar radius are available (cf.~\cite{Bro+94}). 

Sect. 2 of this paper discusses the stars that have been selected as
the most appropriate candidates for observation in the MONS experiment 
(Measuring Oscillations in Nearby Stars) on board the R\o{}mer satellite 
which is an element in the Danish small satellite program (\cite{JCD02}). 
Relevant stellar parameters from the literature or derived in the paper are 
presented here. Sect. 3 presents a set of selection criteria which can 
easily be generalized for other missions with different limiting sensitivities. 
These selection criteria are intended to maximize the effectiveness of any 
oscillation frequencies obtained in constraining stellar structure.  
Sect. 4 presents some results and discussion.


\section{The short-list of stars for MONS}

For the MONS mission a short-list of stars has been decided upon from which a
subset of around 20 stars is to be selected for observation by the main
camera. In making this list, a variety of criteria were used in order to
obtain a set of stars which are representative of several classes or groups
corresponding to the interests of a large international community. 
All of these stars are solar-like in the sense that they are expected
to exhibit oscillations driven stochastically by convection in a
surface convection zone. It is desirable to have stars representative
of several subgroups within this class~: close solar analogues in mass
as well as somewhat lower and higher mass stars (at various stages of
evolution), some stars at lower and higher metallicity than the Sun,
some stars that on the basis of the level of their magnetic activity
are expected to rotate rather more rapidly than the Sun and finally a
few stars at higher effective temperature. 

However it is useful to consider whether it is possible to create a
larger and possibly more uniform sample of stars using quantitative
criteria. Such a list would include most of the stars listed above but
could serve as resource for other asteroseismic missions and observing
campaigns as well as for contingency planning for the MONS mission itself. 
In order to do this it is first necessary to obtain the relevant stellar
parameters. The luminosity $L$ in units of the solar value $L_\odot$
can be obtained from measurable quantities using the following equation~:
\begin{equation}
\begin{array}{rl}
\log {\displaystyle L\over\displaystyle L_\odot} = 4.0+
0.4 M_{{\rm bol},\odot}&-2.0 \log {\pi [{\rm mas}]}\\ 
 -0.4&( V -A_V + BC(V))\,. \\ 
\end{array}
\end{equation}
Here and throughout the paper logarithms are taken to base 10. The
parallax $\pi$ is obtained from the Hipparcos (\cite{HipCat})
satellite in all cases except for a few binary systems. For these the
value is used that is obtained by S\"o{}derhjelm (1999) which takes
into account corrections due to the relative motion of the two
components in their orbit. The Johnson V magnitude is obtained from
the Catalogue of Photometric Data (\cite{GCPD97}). For most stars in
the MONS short-list there are high precision determinations of the
stellar parameters~: effective temperature, surface gravity, and
metallicity $\left(T_{\rm eff}, \log g, [{\rm Fe/H}]\right)$ directly
from spectroscopy. For the remaining stars spectroscopy is being
obtained and will be reported elsewhere (\cite{Bruntt}). With these
values it is possible to obtain colours $(B-V)$, extinction $A_V$ and
bolometric corrections $BC(V)$ on $V$ using the BaSeL library of model
atmospheres (\cite{Lej+98}). $M_{{\rm bol},\odot}$ is the solar
bolometric absolute magnitude.

In order to be fully consistent, in principle it would be necessary to
re-reduce the original spectra or obtain new ones and perform fitting for 
$\left(T_{\rm eff}, \log g, [{\rm Fe/H}]\right)$ using the same model 
atmosphere code that would also be used to calculate the colour and bolometric 
corrections. In practice this is not feasible for a large number of stars
and therefore it is assumed that the uncertainties on the values for 
$\left(T_{\rm eff}, \log g, [{\rm Fe/H}]\right)$ quoted by the various authors
are sufficiently large to encompass any systematic effects between the
various codes currently in use for spectral line fitting for these parameters.
One possible systematic effect is changes in colours and bolometric corrections
that can arise due to different assumptions concerning properties of
convection. In order to asses the magnitude of such effects the
bolometric corrections and colours were compared with those resulting
from using model atmospheres of Bessel et al. (1998) with and without
overshoot. For all stars in the MONS short-list the differences between the two
treatments and between that grid and the BaSeL grid of model
atmospheres were smaller than any differences introduced due to the
uncertainty in the measurements of $\left(T_{\rm eff}, \log g, [{\rm
Fe/H}]\right)$. Of course very different treatments of convection such
as proposed by Canuto \& Dubovikov (1998), or 3-D simulations 
(cf. Asplund et al., 1999) might well introduce substantial changes both in
the estimation of $\left(T_{\rm eff}, \log g, [{\rm Fe/H}]\right)$ as well as
in $BC(V)$ and $(B-V)_0$, but as yet such information is not available for 
large numbers of stars.

   \begin{table*}
   \centering
      \caption[]{Properties of the short-listed stars for the MONS main camera 
      ordered by Hipparcos number. Col. 1: Hipparcos no., 2: name, 3,
      4: spectral type, source, 5,6: known multiplicity, source,
      7,8,9: parallax, $1\sigma$, source, 10,11,12 : effective 
      temperature, $1\sigma$, source, 13,14 : surface gravity, $1\sigma$,
      source in Col. 12. 15,16: $[{\rm Fe/H}]-[{\rm
      Fe/H}]_\odot$, $1\sigma$, source in Col. 12.}
         \label{MONSMaTaI}
\begin{tabular}{rclcccrrrrrrrrrr}
\noalign{\hrule}
 HIP & name &\multispan{2}{\enspace Spec.type$^{\mathrm{a}}$}&\multispan{2}{\enspace Mult.$^{\mathrm{a,b}}$}&$\pi$&$\sigma_\pi$&ref.$^{\mathrm{a}}$
&$T_{\rm eff}$ &$\sigma_T$&ref.$^{\mathrm{a}}$&$\log g$&$\sigma_g$&$[{\rm Fe/H}]$&$\sigma_{[{\rm Fe/H}]}$\\
 &&&&&&\multispan{2}{\hfil [mas]\hfil}& &\multispan{2}{\hfil [K]\hfil}&
&\multispan{2}{\hfil $[{\rm cm/s}^2]$\hfil}&\\
\noalign{\hrule}
  2021&\object{$\beta$ Hyi}   &G2IV  &1& & &133.78&0.51&2&5860& 70& 9&4.05&0.30&-0.11&0.07 \\
  3821&\object{$\eta$ Cas}    &F9V   &1&M&4&167.99&0.62&2&5848&100&11&4.40&0.05&-0.27&0.03 \\
  7513&\object{$\upsilon$ And}&F8V   &1&M&4& 74.25&0.72&2&6135& 61& 8&4.08&0.18& 0.11&0.09 \\
  8102&\object{$\tau$ Cet}    &G8V   &1&B&4&274.17&0.80&2&5264&100&11&4.36&0.05&-0.50&0.03 \\
  8796&\object{$\alpha$ Tri}  &F6IV  &1&M&4& 50.87&0.82&2&6288&   &16&3.91&    & 0.00&     \\
 12777&\object{$\theta$ Per}  &F8V   &1&M&4& 89.03&0.79&2&6248& 80&20&4.20&0.10&-0.01&0.07 \\
 14632&\object{$\iota$ Per}   &G0V   &1&B&4& 94.93&0.67&2&5989& 61& 8&4.19&0.21& 0.16&0.09 \\
 14879&\object{$\alpha$ For}  &F8V   &1&B&4& 70.86&0.67&2&6000&   &17&    &    &-0.35&0.07 \\
 15457&\object{$\kappa^1$ Cet}&G5Vv  &1&M&4&109.18&0.78&2&5576& 61& 8&4.41&0.21& 0.03&0.09 \\
 16537&\object{$\epsilon$ Eri}&K2V   &1&B&4&310.75&0.85&2&5052&100&11&4.57&0.05&-0.15&0.03 \\
 17378&\object{$\delta$ Eri}  &K0IVe &1& & &110.58&0.88&2&4884& 61& 8&3.40&0.18&-0.11&0.09 \\
 19893&\object{$\gamma$ Dor}  &F4III &1& & & 49.26&0.50&2&7300&   &21&4.2 &    &     &     \\
 22449&\object{$\pi^3$ Ori}   &F6V   &1&B&4&124.60&0.95&2&6482& 61& 8&4.35&0.18& 0.05&0.09 \\
 27072&\object{$\gamma$ Lep A}&F6V   &1&M&4&111.49&0.60&2&6302& 61& 8&4.26&0.18&-0.05&0.09 \\
 27913&\object{$\chi^1$ Ori}  &G0V   &1&B&1&115.43&1.08&2&5869& 61& 8&4.45&0.18&-0.01&0.09 \\
 32362&\object{$\xi$ Gem}     &F5III &1& & & 57.02&0.83&2&6464&100&23&3.81&0.02& 0.00&0.01 \\
 37279&\object{$\alpha$ CMi} A&F5IV-V&1&M&4&285.93&0.88&2&6500& 80&10&4.04&0.10& 0.00&0.05 \\
 50954&\object{HR 4102}&F2IV  &1& & & 61.67&0.49&2&7320&260&24&    &    &     &     \\
 55642&\object{$\iota$ Leo}   &F4IV  &1&B&4& 42.6\enspace&1.3\enspace&2&6739& &16&3.98&   & 0.06&    \\
 57757&\object{$\beta$ Vir}   &F9V   &1&B&4& 91.74&0.77&2&6109&100&11&4.20&0.05& 0.17&0.03 \\
 67927&\object{$\eta$ Boo}    &G0IV  &1&B&4& 88.17&0.75&2&6003& 61& 8&3.62&0.18& 0.25&0.09 \\
 70497&\object{$\theta$ Boo}  &F7V   &1&B&4& 68.63&0.56&2&6227& 61& 8&3.84&0.18 &-0.27&0.09\\
 71681&\object{$\alpha$ Cen B}&K1V   &1&B&4&747.1\enspace&1.2\enspace&26&5255&50& 7&4.51&0.08& 0.24&0.03\\
 71683&\object{$\alpha$ Cen A}&G2V   &1&B&4&747.1\enspace&1.2\enspace&26&5830&30& 7&4.34&0.05& 0.25&0.02\\
 71957&\object{$\mu$ Vir}     &F2III &1& & & 53.54&0.95&2&7140&160&22&    &    &     &     \\
 76976&\object{HD 140283}&sdF3  &3& & & 17.44&0.97&2&5687&100&11&3.55&0.05&-2.53&0.03 \\
 77257&\object{$\lambda$ Ser} &G0V   &1&B&5& 85.08&0.80&2&5915&100&11&4.10&0.05&-0.01&0.03 \\
 78072&\object{$\gamma$ Ser}  &F6V   &1&M&4& 89.92&0.72&2&6249& 61& 8&4.16&0.18&-0.15&0.09 \\
 81693&\object{$\zeta$ Her}   &G0IV  &1&B&4& 93.7\enspace&0.60&26&5825&   &14&3.80&  &0.00& \\
 84405&\object{36 Oph A}      &K0V   &1&M&4&167.08&1.07&2&5143&   &19&4.60&    &-0.39&     \\
 86974&\object{$\mu$ Her}     &G5IV  &1&M&4&119.05&0.62&2&5411&100&11&3.87&0.05& 0.16&0.03 \\
 88601&\object{70 Oph A}      &K0V   &1&B&4&195.7\enspace&0.9\enspace&26&5260& &18&5.00&    &-0.25&0.15\\
 89937&\object{$\chi$ Dra}    &F7V   &1&B&4&124.37&0.52&26&6008& 61& 8&4.36&0.18&-0.33&0.10\\
 99240&\object{$\delta$ Pav}  &G6-8IV&1& & &163.73&0.65&2&5538&   &13&3.80&    & 0.28&     \\
102485&\object{$\psi$ Cap}    &F4V   &1& & & 68.16&0.91&2&6632&   &15&4.50&    &-0.11&     \\
104043&\object{$\alpha$ Oct}  &F4III&25&B&6& 22.07&0.57&2&    &   &  &    &    &     &     \\
105858&\object{$\gamma$ Pav}  &F6V   &1& & &108.50&0.59&2&6139& 70& 9&4.34&0.30&-0.67&0.07 \\
109176&\object{$\iota$ Peg A} &F5V   &1&B&4& 85.06&0.71&2&6413& 61& 8&4.16&0.18& 0.00&0.09 \\
110618&\object{$\nu$ Ind}     &A3V  &1&B&4& 34.60&0.60&2&5381&   &12&3.43&    &-1.34&0.08 \\
114996&\object{$\gamma$ Tuc}  &F1III &1& & & 45.40&0.61&2&6541&   &16&3.88&    &-0.27&     \\
\noalign{\hrule}
\end{tabular}
\begin{list}{}{}
\item[$^{\mathrm{a}}$] Ref$^{\rm s}$: 1=Hoffleit\&Warren Jr., Bright Star Cat. 
(1991), 2=The Hipparcos and Tycho Cat. (1997), 3=Cayrel de Strobel et al., 
Cat. of [{\rm Fe/H}] of F, G, K stars (2001), 4=Worley et al., The Washington 
Visual Double Star Cat. (1996), 5=Abt\& Levy (1976), 6=Halbwachs (1981), 
7=Neuforge et al. (1997), 8=Cenarro et al. (2001), 9=Castro et al. (1999),
10=Mashonkina\&Gehren (2000), 11=Soubiran et al. (1998), 12=Gratton et al. 
(2000), 13=Abia et al. (1988), 14=Cunha et al. (2000), 15=Boesgaard\&
Friel (1990), 16=Balachandran (1990), 17=Favata et al. (1997), 18=Zboril\&
Byrne (1998), 19=Cayrel de Strobel et al. (1989), 20=Fuhrmann (1998), 
21=Balona et al. (1994), 22=Malagnini\&Morossi (1990), 23=Lebre et al. (1999), 
24=Sokolov (1995), 25=Buscombe\&Morris (1960), 26=S\"o{}derhjelm (1999).
\item[$^{\mathrm{b}}$] M indicates the star is known to have multiple nearby 
companions which may be physically associated, B indicates the star is known to 
have one (possibly physical) companion, otherwise there is no entry.
\end{list}
   \end{table*}
   \begin{table*}
   \centering
      \caption[]{Further properties of MONS short-listed stars. Col. 1: Hipparcos 
       no., Cols. 2,3,4: $v\sin i$, $1\sigma$ error, source, 5,6:
      Johnson V magnitude, $1\sigma$ error (\cite{GCPD97}), 7,8: Johnson B-V
      colour, $1\sigma$ error (\cite{GCPD97}), 9,10: Visual extinction and
      uncertainty, Cols. 11,12: Luminosity and uncertainty, Cols.
      13,14: Luminosity and uncertainty when setting $A_V=\sigma_A=0$.}
         \label{MONSMaTaII}
\begin{tabular}{rrrrrrllrrrrrl}
\noalign{\hrule}
 HIP &$v\sin i$&$\sigma_{\rm v}$&
ref.$^{\mathrm{a}}$&V&$\sigma_{\rm V}$&B-V
&$\sigma_{\rm B-V}$&$A_V$&$\sigma_A$&$L$ &$\sigma_L$&$L_m$ &$\sigma_L$ \\
 &\multispan{2}{\hfil [km/s]\hfil}&&\multispan{2}{\hfil $[{\rm mag}]$\hfil}&
\multispan{2}{\hfil $[{\rm mag}]$\hfil}&\multispan{2}{\hfil $[{\rm mag}]$\hfil}&
\multispan{2}{\hfil $[L_\odot]$\hfil}&
\multispan{2}{\hfil $[L_\odot]$\hfil}\\
\noalign{\hrule}
  2021&6.0&1.0&29&2.797& 0.006& 0.618&0.005& 0.08&0.07& 4.0\enspace&0.3\enspace&  3.74\enspace&0.05\\
  3821&1.5&0.8&30&3.444& 0.009& 0.572&0.007& 0.00&0.04& 1.32&0.05& 1.323&0.019\\
  7513&9.2&0.7&30&4.086& 0.013& 0.537&0.008& 0.00&0.03& 3.53&0.13& 3.53\enspace&0.09\\
  8102&0.4&0.4&28&3.496& 0.011& 0.727&0.007& 0.00&0.06& 0.52&0.03& 0.523&0.011\\
  8796&90.\enspace&1.0&29&3.414& 0.013& 0.489&0.010& 0.00&0.04&13.8\enspace&0.7\enspace&13.7\enspace\enspace&0.5\\
 12777&8.8&0.6&30&4.107& 0.017& 0.486&0.007& 0.00&0.03& 2.39&0.10& 2.39\enspace&0.06\\
 14632&3.5&0.7&30&4.046& 0.008& 0.593&0.008& 0.01&0.04& 2.30&0.09& 2.28\enspace&0.04\\
 14879&5.2&0.5&35&3.855& 0.018& 0.522&0.011& 0.00&0.04& 5.0\enspace&0.2\enspace& 5.04\enspace&0.15\\
 15457&4.5&0.3&28&4.836& 0.010& 0.679&0.007& 0.00&0.04& 0.89&0.04& 0.892&0.018\\
 16537&1.7&0.3&28&3.726& 0.010& 0.882&0.007& 0.02&0.07& 0.35&0.03& 0.342&0.008\\
 17378&1.0&1.0&29&3.527& 0.012& 0.922&0.007& 0.00&0.05& 3.39&0.19& 3.39\enspace&0.11\\
 19893&50.\enspace&   &36&4.242& 0.004& 0.309&0.005& 0.01&0.03& 6.4\enspace&0.2\enspace& 6.39\enspace&0.15\\
 22449&18.\enspace&1.5&30&3.188& 0.006& 0.450&0.008& 0.00&0.03& 2.77&0.09& 2.77\enspace&0.05\\
 27072&8.7&0.8&32&3.591& 0.013& 0.480&0.012& 0.00&0.03& 2.45&0.08& 2.45\enspace&0.05\\
 27913&9.4&0.4&30&4.401& 0.010& 0.591&0.010& 0.00&0.03& 1.13&0.04& 1.13\enspace&0.02\\
 32362&70.\enspace&7.0&23&3.350& 0.014& 0.437&0.006& 0.00&0.04&11.4\enspace&0.6\enspace&11.4\enspace\enspace&0.4\\
 37279&6.1&1.0&29&0.367& 0.010& 0.421&0.008& 0.00&0.03& 7.0\enspace&0.2\enspace& 7.04\enspace&0.10\\
 50954&50.\enspace&   &38&3.987& 0.010& 0.360&0.006& 0.18&0.13& 6.1\enspace&0.8\enspace& 5.12\enspace&0.11\\
 55642&16.\enspace&1.0&29&3.939& 0.009& 0.409&0.006& 0.02&0.05&11.7\enspace&0.9\enspace&11.5\enspace\enspace&0.7\\
 57757&4.0&0.8&30&3.608& 0.010& 0.551&0.008& 0.00&0.04& 3.59&0.16& 3.59\enspace&0.09\\
 67927&13.\enspace&1.0&29&2.680& 0.009& 0.580&0.009& 0.00&0.03& 9.2\enspace&0.4\enspace& 9.2\enspace\enspace&0.2\\
 70497&29.\enspace&1.0&32&4.049& 0.012& 0.496&0.006& 0.04&0.05& 4.5\enspace&0.2\enspace& 4.32\enspace&0.10\\
 71681&1.1&0.8&28&1.352& 0.010& 0.866&0.026& 0.08&0.09& 0.54&0.05& 0.504&0.008\\
 71683&2.7&0.7&28&0.000& 0.005& 0.651&0.027& 0.01&0.05& 1.58&0.07& 1.556&0.011\\
 71957&46.\enspace&   &34&3.876& 0.012& 0.382&0.005& 0.19&0.09& 9.0\enspace&0.9\enspace& 7.5\enspace\enspace&0.3\\
 76976&   &   &  &7.211& 0.013& 0.490&0.010& 0.00&0.04& 4.3\enspace&0.5\enspace& 4.3\enspace\enspace&0.5\\
 77257&2.4&0.8&30&4.426& 0.009& 0.602&0.007& 0.04&0.06& 2.11&0.13& 2.04\enspace&0.05\\
 78072&9.3&0.7&32&3.842& 0.018& 0.477&0.007& 0.00&0.03& 3.02&0.11& 3.02\enspace&0.08\\
 81693&4.8&1.0&29&2.807& 0.010& 0.644&0.009& 0.09&0.09& 8.2\enspace&0.7\enspace& 7.52\enspace&0.15\\
 84405&0.8&0.9&28&5.05&0.010& 0.960&0.010& 0.45&0.14& 0.52&0.07& 0.344&0.009\\
 86974&1.7&1.0&29&3.417& 0.014& 0.752&0.009& 0.00&0.06& 2.85&0.18& 2.85\enspace&0.08\\
 88601&1.6&0.4&37&4.023& 0.013& 0.864&0.011& 0.26&0.12& 0.80&0.09& 0.631&0.016\\
 89937&2.5&0.4&37&3.571& 0.009& 0.489&0.006& 0.00&0.03& 2.12&0.06& 2.12\enspace&0.03\\
 99240&   &   &  &3.556& 0.011& 0.760&0.002& 0.04&0.08& 1.33&0.10& 1.29\enspace&0.03\\
102485&41.\enspace&   &34&4.137& 0.004& 0.426&0.003& 0.00&0.03& 3.88&0.17& 3.87\enspace&0.11\\
104043&85.\enspace&   &38&5.140& 0.008& 0.488&0.007& 0.03&0.10&15.1\enspace&1.7\enspace&14.8\enspace\enspace&0.9\\
105858&8.\enspace&   &33&4.213& 0.009& 0.484&0.006& 0.01&0.03& 1.56&0.05& 1.55\enspace&0.03\\
109176&6.5&0.8&30&3.768& 0.006& 0.433&0.008& 0.00&0.03& 3.51&0.12& 3.51\enspace&0.07\\
110618&0.0&   &31&5.279& 0.007& 0.657&0.010& 0.05&0.09& 6.7\enspace&0.6\enspace& 6.4\enspace\enspace&0.2\\
114996&80.&5.0&16&3.985& 0.005& 0.400&0.006& 0.00&0.03&10.1\enspace&0.4\enspace&10.1\enspace\enspace&0.3\\
\noalign{\hrule}
\end{tabular}
\begin{list}{}{}
\item[$^{\mathrm{a}}$] Ref$^{\rm s}$ (in addition to Table~1)~: 
28=Saar \& Osten (1997), 29=de Medeiros et al. (1997), 30=Soderblom (1982), 
31=Andersen et al. (1984), 32=Soderblom et al. (1989), 33=Schrijver (1993), 
34=Simon \& Drake (1989), 35=Hale (1994), 36=Soderblom (1983), 37=Gray (1984), 
38=Uesugi \& Fukuda (1992). 
\end{list}
   \end{table*}

A summary of the stellar parameters is presented in Tables 1 and 2. In
a number of cases more than one independent determination of
$\left(T_{\rm eff}, \log g, [{\rm Fe/H}]\right)$ is available in the
literature. However, in many references no error estimates are quoted,
which makes it difficult to asses whether or not the various
determinations are consistent. The values in Tables 1 and 2 are either
from a reference which quotes an error estimate or the most recent
determination available. The bolometric corrections on the V magnitude
$BC(V)$ are obtained by tri-linear interpolation in $T_{\rm eff}$,
$\log g$, and $[{\rm Fe/H}]$ on the tables of Lejeune et al. (\cite{Lej+98})
for the BaSeL library of model atmospheres. In order to be consistent
with these tables, the bolometric magnitude of the Sun $M_{{\rm bol},\odot}$ 
is taken to be 4.746 (\cite{Lej+98}). An estimate of the uncertainty
in the bolometric correction is obtained by treating the error
estimates on $T_{\rm eff}$, $\log g$, and $[{\rm Fe/H}]$ as
independent errors and combining them quadratically. The necessary
partial derivatives of the bolometric correction with respect to each
of these parameters are also obtained from the tables calculated with
the BaSeL library of model atmospheres. For those stars for which no
measurement error for $T_{\rm eff}$, $\log g$, and/or $[{\rm Fe/H}]$
is available the error has been set to $100\ {\rm K}$, $0.2$ and $0.1$
respectively, which in some cases contributes substantially to the
uncertainty in the bolometric correction and the intrinsic colour and 
therefore also to the error estimates for $L$ and $A_V$. 

Since the stars MONS short-listed stars are all nearby, the visual
extinction $A_V$ is expected to be very small. In fact in deriving
relation (\ref{extlaw}) from observations it is often assumed that
there is no extinction within 70~pc of the Sun. Nevertheless in terms
of the uncertainty estimate for the luminosity, in principle the term
must be taken into account. To obtain an estimate of $A_V$ the
standard relation between extinction and reddening is used 
(cf.~\cite{Nec+80})~:
\begin{equation}
A_V = 3.1 \left[(B-V) - (B-V)_0\right]\,,
\label{extlaw}
\end{equation}
where the intrinsic colour $(B-V)_0$ is obtained by
tri-linear interpolation in $T_{\rm eff}$, $\log g$, and $[{\rm Fe/H}]$ on 
the tables of Lejeune et al. (\cite{Lej+98}) for the BaSeL library of model 
atmospheres. The uncertainty  in $(B-V)_0$ is obtained from the BaSeL library 
in the same way as it is for the bolometric correction. This error is then
combined quadratically with the measurement error in $(B-V)$ to obtain an
error estimate on the visual extinction. 

The resulting values for the luminosity $L$ and visual extinction 
$A_V$ are listed in Table~2. In some cases negative values for the visual
extinction are obtained, which can be due to measurement 
uncertainties for $(B-V)$ or the uncertainty in $(B-V)_0$ propagated from
the uncertainties in $\left(T_{\rm eff}, \log g, [{\rm Fe/H}]\right)$.
In such cases the $A_V$ is set to $0$ in Table~2. In all but a few of the
cases where the $A_V$ is set to $0$, the value is between $0$ and 
$-1.5\sigma$. For the exceptions the absolute value of the extinction 
$\vert A_V\vert < 3\sigma$. These exceptions are~:
\begin{itemize}
\item{{\it \object{HIP 37279} ($\alpha$ CMi A):}} it is possible that the
measured $(B-V)$ is affected by the presence of the white dwarf
companion (Procyon B).
\item{{\it \object{HIP 17378} ($\delta$ Eri):}} this star is classified as an
RS CVn type variable (a detached active binary) which could be the
cause for anomalous colours.
\item{{\it \object{HIP 89937} ($\chi$ Dra):}} An astrometric, interferometric,
and spectroscopic binary. If the secondary is also a main sequence
star it should not have a higher $T_{\rm eff}$ in which case it would
not be likely to cause an anomalously blue colour of the combined binary. 
\item{{\it \object{HIP 109176} ($\iota$ Peg A):}} A spectroscopic
binary with a period of $\sim 10\ {\rm d}$. Possibly the companion is
sufficiently near in mass to affect the colour. 
\end{itemize}

   \begin{figure*}
   \sidecaption
   \includegraphics[width=12cm]{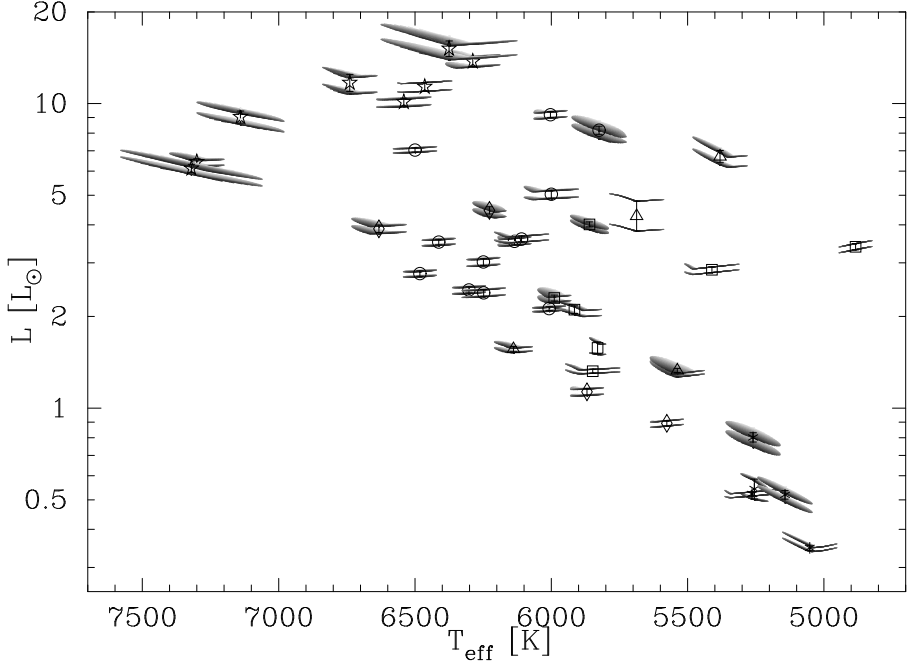}
   \caption{The theoretical HR diagram of the MONS short-listed stars. Symbols
     are : squares: solar mass, asterisks: lower mass, circles: higher mass, 
     triangles: lower and higher metallicity, diamonds: rapid rotator, 
     stars: hotter. For each star is shown the upper and lower
     $1\sigma$ error bound around the luminosity for the appropriate $T_{\rm eff}$ 
     within its $1\sigma$ error. The shading from light to dark
     of each elliptical surface corresponds to $[{\rm Fe/H}]$ running from the
     lowest to the highest value within its $1\sigma$ error bound.}
              \label{FigHR}%
    \end{figure*}
As expected for nearby stars, for all stars in Table~2 but one the value for 
$\vert A_V\vert < 3\sigma(A_V)$. The one exception is 36 Oph A (\object{HIP
84405}) which has a nearly equal mass companion so that for this star the
determination of $A_V$ is probably an overestimate. It is important
for high precision photometry to know whether or not a target star has one or more
close companions, for instance because it is member of a (close) binary 
or multiple system, which is why this is indicated in Table~2. One star for
which binarity is indicated is $\nu$ Ind (\object{HIP 110618}). However there
is some doubt as to the binarity of this star (cf. \cite{LamMcW}) and
evidently the spectral type is not consistent with the spectroscopic
effective temperature.

The uncertainty in the luminosity in Table~2 is obtained by treating each
of the terms in Eq. (2) as a statistically independent source of errors
and combining the variances appropriately. Although 
the number obtained in this way does provide a measure of the range of 
luminosities consistent with the data, it is misleading in that the error
distribution is not Gaussian since there is a correlation between e.g.
the effective temperature and the bolometric correction on V. In the
theoretical HR diagram Fig.~\ref{FigHR} account is taken of this by 
showing for each star the $1\sigma$ lower and upper bound for the
luminosity obtained for $T_{\rm eff}$ and $[{\rm Fe/H}]$ within their
combined $1\sigma$ error bounds. In practice in all cases the largest
remaining source of error is due to the uncertainty in $A_V$ which
arises in most cases primarily from the $1\sigma$ error on the
observed $B-V$ colour. In Table~2 the minimum luminosity $L_m$ and its
uncertainty for each star are also given, calculated as above but
setting both $A_V=0$ and $\sigma_{A}=0$. The uncertainty $\sigma_{L}$
obtained in this way has roughly equal contributions from the errors
in the V magnitude, the parallax, and the bolometric correction.

   \begin{table}
   \centering
      \caption[]{Dynamic masses of some of the MONS short-listed stars.}
         \label{MONSDynM}
\begin{tabular}{rllr}
\noalign{\hrule}
 HIP No. & $M$ & $\sigma_M$ & ref.$^{\mathrm{a}}$ \\
 &\multispan{2}{\hfil $[M_\odot]$\hfil}& \\
\noalign{\hrule}
  3821&0.91 & 0.05  &1\\
 27913&1.02 & 0.08  &7\\
 37279&1.497& 0.037 &2\\
 55642&1.7  & 0.2   &5\\
 71681&0.934& 0.0061&3\\
 71683&1.105& 0.0070&3\\
 81693&1.25 & 0.05  &1\\
 88601&0.90 & 0.074 &4\\
 89937&0.98 & 0.03  &5\\
109176&1.31 & 0.02  &6\\ 
\noalign{\hrule}
\end{tabular}
\begin{list}{}{}
\item[$^{\mathrm{a}}$] Ref$^{\rm s}$: 1=Harmanec(1988), 2=Girard et~al.(2000), 
3=Pourbaix et~al.(2002), 4=Pourbaix(2000), 5=S\"oderhjelm(1999),
6=Fekel \& Tomkin(1983), 7=K\"onig et~al.(2002).
\end{list}
   \end{table}

   \begin{figure}
   \centering
   \includegraphics[width=88mm]{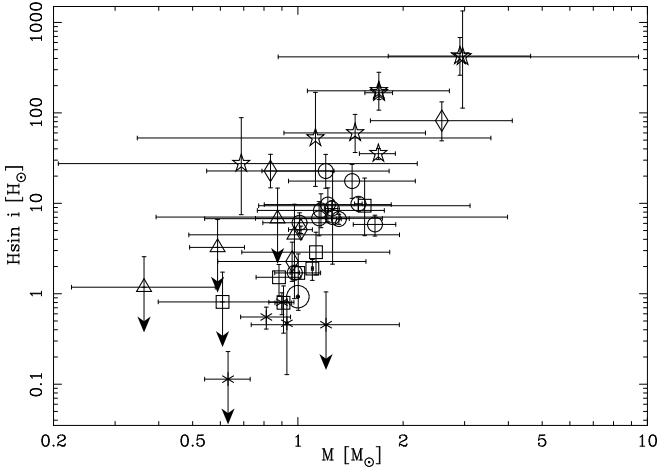}
   \caption{Angular momentum versus mass for the MONS short-listed  
     stars. Symbols as in Fig. 1. The symbol $\odot$ is for the Sun,
     located away from unity due to differential rotation as explained 
     in the text.}
              \label{FigHvsM}%
    \end{figure}

From Fig.~\ref{FigHR} it is seen that the sample covers a range in masses and
ages which is essential for the purposes of testing stellar evolution theory.
In Fig.~\ref{FigHR} each star is represented by two elliptical
surfaces, in some cases with a `kink', connected by a vertical bar at
the $T_{\rm eff}$ listed in Table~1. The bolometric correction depends
on $T_{\rm eff}$, and $[{\rm Fe/H}]$ and to a lesser extent on $\log
g$, which means that the $1\sigma$ range for the luminosity is not
identical for every combination of $T_{\rm eff}$ and $[{\rm Fe/H}]$ 
within their combined $1\sigma$ error ellipse. For each
star the upper ellipse represents the $1\sigma$ upper limit for the
luminosity and the lower ellipse represents the $1\sigma$ lower
limit. The shading of each ellipse corresponds to the $[{\rm Fe/H}]$
running from the central $-1\sigma$ to $+1\sigma$ value. For some
stars the latter source of uncertainty is so small that the ellipses 
reduce to lines. The `kinks' can occur because the intrinsic colour
of the star becomes bluer, $(B-V)_0$ smaller, as the effective 
temperature is higher. From Eq. (\ref{extlaw}) it can be seen 
that for $T_{\rm eff}$ higher than a certain value $A_V$ becomes
positive and therefore the inferred luminosity after correcting for
this extinction increases towards higher $T_{\rm eff}$, whereas for
lower $T_{\rm eff}$ $A_V$ remains fixed at $0$ since a negative
extinction is unphysical. 

As is discussed in Sect. 1, the aims of asteroseismic campaigns include 
obtaining masses and rotation rates of stars with higher precision than 
currently available.
It is therefore of interest to show diagrams for the mass and the angular 
momentum of these stars which can be determined directly, even if
only with low precision and up to a factor $\sin i$ in the unknown inclination 
angle, from the data listed in Tables 1 and 2. In Fig.~\ref{FigHvsM}
the angular momentum is shown versus the stellar mass.
The (spectroscopic) mass of the stars can be estimated using~:
\begin{equation}
\log {M\over M_\odot} = \log {g\over g_\odot} + \log {L\over L_\odot} - 
4\log{T_{\rm eff}\over T_{{\rm eff},\odot}}\,.
\end{equation}
Similarly the (spectroscopic) angular momentum $H$ can be estimated as follows~:
\begin{equation}
\begin{array}{rl}
\log {\displaystyle H\sin i\over\displaystyle  H_\odot} = \log 
{\displaystyle g\over\displaystyle  g_\odot} + {\displaystyle 
3\over\displaystyle 2}\log &{\displaystyle L\over\displaystyle  L_\odot} 
- 6\log {\displaystyle T_{\rm eff}\over\displaystyle  T_{{\rm eff},\odot}} \\
&+ \log {\displaystyle I_M} + \log {\displaystyle v\sin i
\over\displaystyle v_\odot}\,. \\
\end{array}
\label{specH}
\end{equation}
However, for a few stars a dynamic mass $M_{\rm dyn}$ is available
which is preferable to use since it is always more accurate. In this
case the appropriate equation for $H$ is~:
\begin{equation}
\begin{array}{rl}
\log {\displaystyle H\sin i\over\displaystyle H_\odot} = \log 
{\displaystyle M_{\rm dyn}\over\displaystyle M_\odot} + {\displaystyle 1\over
\displaystyle 2} \log &{\displaystyle L\over\displaystyle L_\odot} 
- 2\log {\displaystyle T_{\rm eff}\over\displaystyle  T_{{\rm eff},\odot}} \\
&+ \log {\displaystyle I_M} + \log {\displaystyle v\sin i
\over\displaystyle v_\odot}\,. \\
\end{array}
\label{dynH}
\end{equation}
Those stars in the MONS short-list for which dynamic masses are available are
summarized in Table~3. 

In Eqs. (\ref{specH}) and (\ref{dynH}) the term $I_M$ is the
scaled moment of inertia of the star~:
\begin{equation}
I_M = {8\pi\over 3 M R^2} \int\limits_{0}^{R} \rho r^4 {\rm d}r
\label{inermom}
\end{equation}
The total angular momentum of a star is an integral over the stellar
interior of the angular momentum per unit mass, i.e. a density
weighted average of the rotation rate $\Omega(r,\theta)$. If the stars
were to rotate uniformly the above equations would hold exactly. For
the Sun the true angular momentum is known to be $H_\odot = 190 \pm 1.5\ \times
10^{39}\ {\rm kg\ m^2\ s^{-1}}$ (\cite{Pijp98}) which is determined 
helioseismically taking into account the full radial and latitudinal 
dependence of the rotation rate $\Omega_\odot (r,\theta)$. Using this
value in Eq. (\ref{dynH}), a value for $I_M$ calculated from a standard
solar model (cf. \cite{JCD+96}), and using the observed equatorial surface
rotation velocity $v_\odot = 1.93\ {\rm km/s}$ the deviation from a
value of $1$ is less than 2\%, because the deviations from uniform
rotation in the Sun are small. However, even if the interior density
of a star can be calculated from models, the interior rotation rate of
stars other than the Sun is still unknown. Therefore in producing
Fig.~\ref{FigHvsM}, using Eq.~(\ref{specH}) or
(\ref{dynH}), the value of $I_M$ is kept fixed at $I_{M\,\odot}$. 
For stellar models the value of $I_M/I_{M\,\odot}$ decreases with
increasing mass and with increasing age on the main sequence. In the
mass range of 0.85 to 4.0 solar masses, and with ages spanning the
main sequence lifetime, this ratio lies between 0.5 and 2.0. Including
this in Fig.~\ref{FigHvsM} would have the effect of making the trend
slightly less steep. In order to be able to show error bars on the
angular momentum estimated using Eq.~(\ref{specH}) or (\ref{dynH}) for
those stars in Table~2 for which there is no error estimate for $v\sin
i$, $\sigma_v$ is set to $5.0\ {\rm km/s}$. For the stars for which
no determination of $v\sin i$ is available $v\sin i$ and $\sigma_v$
are both set to $5.0\ {\rm km/s}$ as well. In the latter case the
value is an upper limit, which is indicated by downward arrows in
Fig.~\ref{FigHvsM}. The `rapid rotators' do not appear to stand out
significantly from the general trend in Fig.~\ref{FigHvsM}. Clearly 
for any significant determination of a dependence of angular momentum
on mass or age, a much higher precision for both mass and angular
momentum is required. A precision similar to that of the dynamical
masses in Table~3 is realistic to expect from asteroseismic data to be
collected by the various satellite missions.

\section{The selection criteria}
  
The focus of the MONS asteroseismic mission is on solar-like stars for
two reasons~: firstly it is important for the purposes of mode identification
as well as for the quality of the asteroseismic inferences to be made that 
many modes of pulsation be excited
simultaneously. The highest probability for this to occur is in stars where
the oscillations are excited by convection such as the Sun, because in
principle this can excite all possible modes contrary to other excitation 
mechanisms. The second reason for choosing stars not too dissimilar
from the Sun is that asteroseismic inference is still in its
infancy. The likelihood of success in interpreting asteroseismic data
is higher for stars that are in most respects close analogues of the
Sun. The MONS short-listed stars span the range of $T_{\rm eff}$ for
which this can be said to be the case. Spectroscopically obtained
values (and their uncertainties !) for $\left(T_{\rm eff}, \log g,
[{\rm Fe/H}]\right)$ are considered to be essential in the
determination of high quality stellar parameters, rather than relying
on colour-calibrations. One of the criteria to be used in forming a
sample of stars is that such spectroscopy has been done. Unfortunately
with currently available data this leads to a biased sample of
stars. It would be desirable to be able to have spectroscopy for a
volume-limited sample of stars but this requires a considerable
investment both observationally and in terms of data analysis and 
interpretation. 

The driving factor for the selection criteria is to obtain useful
estimates of the bolometric luminosity from the observed data,
since it is this parameter together with the effective temperature
which are currently used as constraints for stellar evolution models.
In order to be able to do photometry at the ppm level with a
spaceborn telescope with an collecting area of $\sim 600\ {\rm cm}^2$
(i.e. MONS), the stars must be bright. However, the space density of 
metal poor stars is quite low which 
means that any sample which does not differentiate in metallicity will
contain very few metal-poor stars. In order to obtain a reasonably large
sample of Pop. I and II stars, a separate magnitude limit for each group
appears desirable. At the same luminosity a fainter magnitude limit 
corresponds to larger distances. Two different criteria for the parallax
are therefore proposed as well. In the MONS short-list of potential targets, 
the limiting factor on the precision of the luminosity is the uncertainty in 
$A_V$ which is typically around $0.06$. Setting the estimated error due to 
the error in the parallax to be maximally the same value implies that 
$\pi/\sigma(\pi) >36.2$. The best parallax determinations from the Hipparcos 
catalogue have $\sigma(\pi) = 0.45\ {\rm mas}$ so that no stars with a parallax 
$\pi < 16.2\ {\rm mas}$ would qualify. 

The quantitative selection criteria for suitable targets for the MONS mission
can now be summarized as follows~:
\begin{itemize}
\item{} Select stars for which spectroscopic determinations of
$\left(T_{\rm eff}, \log g, [{\rm Fe/H}]\right)$ are available (eg. from
\cite{CSCat}) and for which $4850\ {\rm K} < T_{\rm eff} < 7350\ {\rm K}$.
\item{} Two separate brightness criteria are proposed~:
Group (a) stars, for which $[{\rm Fe/H}] < -0.6$, pass if their
$V$ magnitude satisfies $V<7.3$. Group (b) stars pass 
if their $V$ magnitude satisfies $V<5.3$. The corresponding noise level 
($4\sigma$ detection limit in amplitude after 30 days observing with a 
duty cycle of 85 \% for the faintest stars in the group) of the photometry 
for the MONS mission is $\sim 13\ {\rm ppm}$ for group (a), and 
$\sim 4.5\ {\rm ppm}$ for group (b) (\cite{KjeBed}).
\item{} The two separate distance criteria for group (a) and (b),
corresponding in space volume to the two separate limiting magnitudes,
are~: the group (a) of stars for which $[{\rm Fe/H}] < -0.6$ pass if
their parallax satisfies $\pi/\sigma(\pi) >14$ (which implies $\pi >
6.4\ {\rm mas}$), and the group (b) of stars pass if their parallax
satisfies $\pi/\sigma(\pi) >36$ (which implies $\pi > 16.2\ {\rm mas}$). 
\end{itemize}
By selection from the Hipparcos catalogue (\cite{HipCat}) there are
407 stars which pass the criteria $V<5.3$ and $\pi/\sigma(\pi) >36$
(group (b)), but there is good spectroscopy currently available in the
literature for only 196 of them. For the selection of stars for group
(a) ($V<7.3$ and $\pi/\sigma(\pi) >14$) there are 2428 additional
candidates from the Hipparcos catalogue but in order to apply the
metallicity criterion good spectroscopy must be available, so it is
unclear how many of these stars would in fact qualify. The number of
stars for which good spectroscopy is available which pass all the criteria
for group (a), including the metallicity criterion, currently stands
at 36. 

It should be noted that still not all stars selected in this way are
suitable for performing asteroseismology. The satellite missions are
dependent on doing high precision time-series photometry, and the presence 
of possibly variable background stars within arcmin of the target can
have serious detrimental effects on the signal-to-noise ratio. The
stellar census from surveys is usually incomplete very near to bright 
stars because of saturation effects. Therefore dedicated imaging of the near
field of potential targets has been or is being undertaken for the COROT
and MONS missions. Furthermore, for a mission such as COROT which
will select a very small set of fields in order to obtain very long 
time-series, the selected fields should preferably have within them
a number of unsaturated, bright stars representing as wide a range of $(T_{\rm
eff}, \log g, [Fe/H])$ as possible, which is a more complex
optimization problem. 

\section{Conclusions}
For those stars that are the main targets for the MONS experiment 
on board the planned Danish small satellite R\o{}mer, the luminosity is
rederived using the BaSeL library of model atmospheres (\cite{Lej+98}). 
It is shown that even for these stars, the brightest in their 
respective classes, the uncertainty in their fundamental
parameters is still substantial in some cases and it is worthwhile
obtaining, or re-analysing, high quality spectra. It is also clear
that for single stars the direct constraint on stellar masses from
spectroscopy is weak. The constraints on angular momenta of stars from
spectroscopy are weak as well since there is little knowledge of
the internal rotation of stars and there is always the $\sin i$
ambiguity. For binaries one might assume that the rotation axes are
perpendicular to the orbital plane, but this implies a certain
exchange of angular momentum between the star and the orbit and 
therefore a bias is introduced. This implies that the angular
momentum distribution and evolution of field stars in the galaxy is poorly
constrained. It is here that asteroseismology can provide considerable
improvement through determination of stellar masses (cf. \cite{JCD93})
and also stellar angular momenta (Pijpers, 2003).

Using the present set as a template, a set of objective sample
selection criteria for stars of spectral type F, G, and K can be
defined. It is clear that there are several hundred stars for which
asteroseismology from space could potentially provide strong
constraints on stellar structure and evolution modelling. However,
with currently available spectroscopic data the full potential can be
exploited for less than half of those stars. In preparation for the
ESA space mission Eddington it is desirable to carry out extensive
ground-based programs of high-resolution high S/N spectroscopy for
these stars, and to do detailed stellar atmosphere modelling in order
to obtain precise fundamental parameters.

\begin{acknowledgements}
The author thanks J\o{}rgen Christensen-Dalsgaard, Hans Bruntt, 
Hans Kjeldsen, and Teresa Teixeira for many useful comments and
discussions, and the referee R. Garrido for suggestions improving the
presentation of the paper. The author also thanks the Theore\-tical
Astro\-physics Centre, a collaborative centre between Copenhagen
University and Aarhus University funded by the Danish Research
Foundation for support of this work. This research has made use of the
VizieR catalogue access tool, CDS, Strasbourg, France.
\end{acknowledgements}

\end{document}